# HYDRODYNAMIC SIMULATIONS OF GALAXY FORMATION

Anne A. Thoul
Institute for Advanced Study, Princeton, NJ 08540.
thoul@guinness.ias.edu

## ABSTRACT

We have developed an accurate, one-dimensional, spherically symmetric, Lagrangian hydrodynamics/gravity code, designed to study the effects of radiative cooling and photo-ionization on the formation of protogalaxies. We examine the ability of collapsing perturbations to cool within the age of the universe. In contrast to some studies based on order-of-magnitude estimates, we find that cooling arguments alone cannot explain the sharp upper cutoff observed in the galaxy luminosity function. We also look at the effect of a photoionizing background on the formation of low-mass galaxies.

## 1. INTRODUCTION

There have been two quite different approaches to the study of the effects of gas dynamics in galaxy formation. One uses simple analytic estimates (Binney 1977; Silk 1977; and Rees & Ostriker 1977): typically, one computes the characteristic density and virial temperature of a dark halo assuming a spherical collapse model, then asks whether gas at this density and temperature can cool within a dynamical time, or within a Hubble time. Combined with extended versions of the Press-Schechter (1974) formalism, these methods can yield detailed predictions for properties and evolution of the galaxy population (White & Rees 1978; White & Frenk 1991; Kauffman, White & Guiderdoni 1993; Cole et al. 1994). The second approach is to incorporate gas dynamics directly into three-dimensional numerical simulations (e.g. Katz & Gunn 1991; Cen & Ostriker 1992, 1993; Katz, Hernquist & Weinberg 1992; Evrard, Summers & Davis 1994; Steinmetz & Muller 1994).

In the work reported here (see Thoul & Weinberg 1995a,b for full versions), we take an intermediate path, modeling the collapse of individual perturbations with a *one-dimensional*, Lagrangian, gravity/hydro code. The code evolves a mixture of gas and collisionless dark matter, elements of which are represented by concentric, spherical shells. The gas responds to gravity and pressure forces; it can be heated by adiabatic compression, by shocks, and by energy input from a photo-ionizing background, and it can cool by a variety of atomic radiative processes. The collisionless dark matter responds only to gravitational forces. We adopt initial conditions appropriate to Gaussian random fluctuations, as might be produced by inflation in the early universe.

The geometry in our calculations is idealized, and one must therefore take care to keep their limitations in mind. Nonetheless, they provide a valuable complement to their more elaborate, 3-dimensional cousins because of their high resolution, their speed, and their relative simplicity. Three-dimensional hydrodynamic simulations of galaxy formation suffer from limited spatial resolution and mass resolution, making it difficult to separate genuine physical results from numerical artifacts.

Here we apply the code to two of the basic questions of galaxy formation: what causes the abrupt cutoff at the upper end of the galaxy luminosity function, and does photoionization by the UV background prevent the formation of dwarf galaxies?

## 2. MODEL

The gaseous component is described by the fluid equations for a perfect gas. The collisionless component is described by the equation of motion $dv_d/dt = -M(r_d)/r_d^2$, where $r_d$ and $v_d$ are the radii and velocities of the collisionless mass shells, and $M(r_d)$ is the total (baryonic and dark matter) mass inside radius $r_d$. We use the standard, second-order accurate, Lagrangian finite-difference scheme (Bowers & Wilson 1991).

The radiative cooling is calculated for a gas of primordial composition, 76% hydrogen and 24% helium by mass. We compute the abundances of ionic species as a function of density and temperature by assuming that the gas is in ionization equilibrium with a spatially uniform background of UV radiation. In other words, we choose the abundances so that the rate at which each species is depopulated by photo-ionization, collisional ionization, or recombination to a less ionized state is equal to the rate at which it is populated by recombination from a more ionized state or by photo-ionization or collisional ionization of a less ionized state. The intensity and spectrum of the UV background are specified as a function of time based on theoretical models or observational constraints. Here, we adopt a uniform and constant UV background given by $J_\gamma = J_0(\nu_L/\nu)^\alpha$, where $\nu_L$ is the Lyman limit and $J_0 = J_{21} \times 10^{-21}$ ergs s$^{-1}$ Hz$^{-1}$ cm$^{-2}$ sr$^{-1}$. For the physical problems that we study here, the timescales for reaching ionization equilibrium are much shorter than the other timescales of interest, so our equilibrium assumption should be an excellent approximation. Photo-ionization can affect the behavior of lower mass perturbations because it eliminates the neutral hydrogen and singly ionized helium that dominate cooling at low temperatures, and because the residual energy of the photo-electrons heats low-density gas to $T \sim 10^4$K.

As initial conditions, we adopt the average density profile around a $2\sigma$ peak in a Gaussian random density field (Bardeen et al. 1986). These initial conditions have two free parameters, which we choose to be the filter mass $M_f$, i.e., the mass contained within a sphere of radius $2R_f$ where $R_f$ is the filter radius, and the collapse redshift $z_c$, defined to be the redshift at which the $r = 2R_f$ shell would collapse to $r = 0$ in the absence of pressure.

## 3. RESULTS

We have performed a series of collapse calculations in which we vary the value of the filter mass $M_f$ while keeping the collapse redshift fixed at $z_c = 2$. In these calculations, we use $\Omega_d = 0.9$ and $\Omega_b = 0.1$. The solid points in Figure 1 show the results for $J_{21} = 0$ (i.e., no UV background). We plot the ratio of the mass $M_c$ of gas that cools by $z = 0$ to the mass $M_c(p = 0)$ of gas that would cool and collapse to the center by $z = 0$ in the case of perfectly efficient radiative cooling (zero pressure). We characterize the perturbations by the circular velocity $v_c = (GM_f/0.5r_{ta})^{1/2}$ that would be associated with the shell of interior mass $M_f$ if it collapsed and virialized at half its turnaround radius $r_{ta}$. In the absence of a UV background, the radiative cooling cuts off very sharply at $T \approx 10^4$ K. The solid points in Figure 1 show three distinct regimes: for very low circular velocities, $v_c \lesssim 7$ km/s, the post-shock gas remains below $10^4$K, and there is no cooling at all; at intermediate circular velocities, $7$ km/s $\lesssim v_c \lesssim 30$ km/s, all of the shocked gas cools; and at high circular velocities, $v_c \gtrsim 30$ km/s, only a fraction of the shocked gas cools. However, the transition from the intermediate-velocity regime to the high-velocity regime is not a sharp one. Indeed, in the latter case, the outermost shocked shells remain pressure supported all the way to $z = 0$ and do not cool, but the inner shells of these perturbations collapse earlier, at higher density and lower virial temperature, and these shells are able to cool. Therefore, even though the ratio $M_c/M_c(p=0)$ decreases with increasing $v_c$ in this regime, the decrease is quite slow. In particular, $M_c/M_c(p=0)$ falls much less rapidly than $1/M_c(p=0)$, so the actual amount of cooled gas *increases steadily* with $v_c$. We see that the requirement that gas be able to cool within a Hubble time cannot by itself explain the sharp upper cutoff in the luminous mass of observed galaxies.

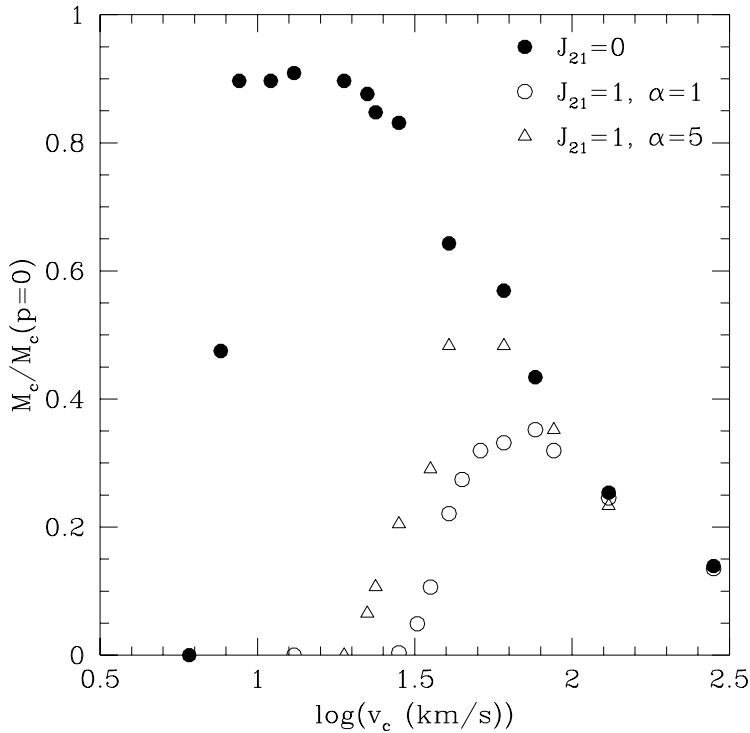

Fig. 1. Ratio of the mass $M_c$ of gas that cools by $z = 0$ to the mass $M_c(p = 0)$ of gas that would cool and collapse by $z = 0$ in the absence of pressure, as a function of the circular velocity $v_c$.

It has been speculated that the presence of a background of UV radiation could inhibit the formation of dwarf galaxies (Efstathiou 1992). We have also performed collapse calculations in the presence of a uniform UV background of amplitude $J_{21} = 1$. Here again, we have kept a fixed $z_c = 2$, $\Omega_d = 0.9$ and $\Omega_b = 0.1$, and we varied the value of the filter mass $M_f$. The open circles in Figure 1 show the results for $\alpha = 1$, a spectrum slope we might expect if the UV radiation is mostly produced by quasars. In this case, we see that the UV background completely suppresses the formation of galaxies with circular velocities $v_c \lesssim 30\,\text{km/s}$. The open triangles show the results for $\alpha = 5$, a softer spectrum that might arise if the UV radiation is mostly produced by star-forming galaxies. Here the UV background suppresses the formation of galaxies with circular velocities $v_c \lesssim 20\,\text{km/s}$. The presence of a photoionizing background considerably affects the amount of gas allowed to cool and form galaxies up to circular velocities of the order of $80\,\text{km/s}$ for $\alpha = 1$ and $50\,\text{km/s}$ for $\alpha = 5$. For larger circular velocities, however, the virial temperature is high, the gas is collisionally ionized, and the cooling is dominated by free-free transitions. Including a photo-ionizing background does not, therefore, alter the results at high circular velocities.

Thus, even though cooling arguments alone cannot explain the sharp upper cutoff in the observed galaxy luminosity function, the presence of the UV background significantly affects the results at the low-mass end. We are presently examining the role of the spectrum and history of the UV background in more detail.

## ACKNOWLEDGEMENTS

This research was supported by the Ambrose Monell Foundation (AT), the W.M. Keck Foundation (DW) and by NSF grant PHY92-45317. We enjoyed the hospitality and stimulating atmosphere of the Aspen Center for Physics during part of this work.